\documentclass[superscriptaddress,aps,prd,showpacs,twocolumn]{revtex4}
\usepackage{graphicx}
\usepackage{amsmath}
\usepackage{bm}
\usepackage{color}

\def\be{\begin{equation}}
\def\ee{\end{equation}}
\def\bea{\begin{eqnarray}}
\def\eea{\end{eqnarray}}

\begin{document}

\title{A Riccati equation based approach to isotropic scalar field cosmologies }
\author{Tiberiu Harko}
\email{t.harko@ucl.ac.uk}
\affiliation{Department of Mathematics, University College London, Gower Street, London
WC1E 6BT, United Kingdom}
\author{Francisco S. N. Lobo}
\email{flobo@cii.fc.ul.pt}
\affiliation{Centro de Astronomia e Astrof\'{\i}sica da Universidade de Lisboa, Campo
Grande, Edific\'{\i}o C8, 1749-016 Lisboa, Portugal}
\author{M. K. Mak}
\email{mkmak@vtc.edu.hk}
\affiliation{Department of Computing and Information Management, Hong Kong Institute of
Vocational Education, Chai Wan, Hong Kong, P. R. China}
\date{\today}

\begin{abstract}
Gravitationally coupled scalar fields $\phi $, distinguished by the choice
of an effective self-interaction potential $V(\phi )$, simulating a
temporarily non-vanishing cosmological term, can generate both inflation and
late time acceleration. In scalar field cosmological models the evolution of
the Hubble function is determined, in terms of the interaction potential, by
a Riccati type equation. In the present work we investigate scalar field
cosmological models that can be obtained as solutions of the Riccati
evolution equation for the Hubble function. Four exact integrability cases
of the field equations are presented, representing classes of general
solutions of the Riccati evolution equation. The solutions correspond to cosmological models in which the Hubble function is proportional to the scalar field potential plus a linearly decreasing function of time, models with the time variation of the  scalar field potential proportional to the potential minus its square, models in which the potential is the sum of an arbitrary function and the square of the function integral, and models in which the potential is the sum of an arbitrary function and the derivative of its square root, respectively. The cosmological
properties of all models are investigated in detail, and it is shown that they can describe the inflationary or the late accelerating phase in the evolution of the Universe. \\
\\
{\bf keywords}: Riccati equation; isotropic scalar field; cosmology 
\end{abstract}

\maketitle



\section{Introduction}


The standard model of cosmology is remarkably successful in accounting for
the observed features of the Universe. However, there remain a number of
fundamental open questions at the foundation of the standard model. In
particular, we lack a fundamental understanding of the acceleration of the
late universe \cite{latetime}. In fact, the standard model of cosmology has
favoured dark energy models, involving time-dependent scalar fields, as
fundamental candidates responsible for the cosmic expansion. Indeed, scalar
fields naturally arise in particle physics, including string theory, and in
addition to this, the underlying dynamics in inflationary models depend
essentially on a single scalar field, with the inflaton rolling in some
underlying potential \cite{1}. A plethora of candidates exist for dark
energy, and we refer the reader to \cite{Copeland:2006wr} for a review.

In a cosmological context, the mathematical properties of the
Friedmann-Robertson-Walker (FRW) models with a scalar field as a matter
source have also been extensively investigated. For instance, in \cite%
{Mus,Sal} a simple way of reducing the system of the gravitational field
equations to a first order equation was proposed, namely, to the
Hamilton-Jacobi-like equation for the Hubble parameter $H$ considered as a
function of the scalar field $\phi $. In fact, a number of integrable
one--scalar spatially flat cosmologies, which play a natural role in the
inflationary scenarios, were studied in \cite{Fre1, Fre2}. Recently, a
general method for the study of scalar field cosmologies, based on the
reduction of the Klein-Gordon equation to a first order non-linear
differential equation, was proposed in \cite{HLM}.

It is interesting to note that the possibility of describing the
cosmological dynamics for a barotropic fluid in terms of a Riccati equation
was discussed in \cite{Far}. For a cosmological fluid satisfying an equation
of state of the form $p=(\gamma -1)\rho $, the Friedmann equations give for
the scale factor the evolution equation in the conformal time $\eta $ is
written as $a^{\prime \prime }/a+(c-1)\left( a^{\prime }/a\right) ^{2}+ck=0$%
, where $c=3\gamma /2-1$, and $k=\pm 1$, and a prime denotes the derivative
with respect to $\eta $. By introducing the transformation $u=a^{\prime }/a$%
, we obtain for $u$ the Riccati type equation $u^{\prime }+cu^{2}+ck=0$,
which is easily integrable. The Riccati equation based study of the
different properties of the isotropic FRW type cosmological models was
performed in \cite{Rosu,10b,10c, 10d,10e}. In fact, the integrability conditions of the
Riccati equation obtained in \cite{MHR,MHR2} allow the integration of the
structure equations of isotropic general relativistic compact objects \cite%
{MHS} in the context of general relativity. Furthermore, the applications of
the Riccati equation to stellar and cosmological models have been
extensively discussed in the literature, and we refer the reader to \cite%
{MH6,MH3,MH4,MH5,U1}. Very recently, using the Chiellini type integrability
condition for the generalized first kind Abel differential equation,
consequently, the new class of exact analytical solution of the Riccati type
equation has been obtained in \cite{U2}.

Therefore, the theoretical investigation of scalar field models is an
essential task in cosmology. It is the purpose of the present paper to
explore alternative approach in solving the cosmological gravitational field
equations in the presence of self-interacting scalar fields, based on the
mathematical analysis of the Riccati type equation that gives the Hubble
function in terms of the scalar field potential. We present several cases of
integrability of this equation, corresponding to specific forms of the
scalar field potential, or of the Hubble function. The physical properties
of the obtained solutions are analyzed in detail.

The present paper is organized as follows. The basic Riccati evolution
equation for scalar field cosmologies with an arbitrary self-interaction
potential is derived in Section \ref{sect2}, and four classes of exact
scalar field solutions are obtained in Section \ref{sect3}. We discuss and
conclude our results in Section \ref{sect7}. Throughout this paper, we use
natural units $c=8\pi G=\hbar =1$, and adopt as our signature for the metric
$\left( +1,-1,-1,-1\right) $.


\section{The Riccati evolution equation for the Hubble function: General
formalism}

\label{sect2}


Consider in the Einstein frame the following Lagrangian density, which
represents a general class of scalar field models, minimally coupled to the
gravitational field,
\begin{equation}
L=\frac{1}{2}\sqrt{\left| g\right| }\left\{ R+ \left[ g^{\mu \nu
}\left( \partial _{\mu }\phi \right) \left( \partial _{\nu }\phi \right)
-2V\left( \phi \right) \right] \right\} \,,
\end{equation}
where $R$ is the curvature scalar, $\phi $ is the scalar field, and $V(\phi )
$ is the self-interaction potential.

Assume a flat FRW scalar field dominated Universe given by the following the
line element
\begin{equation}
ds^{2}=dt^{2}-a^{2}(t)\left( dx^{2}+dy^{2}+dz^{2}\right) ,
\end{equation}%
where $a$ is the scale factor. Thus, the evolution of a cosmological model
is governed by the system of the field equations
\begin{eqnarray}
3H^{2} &=&\rho _{\phi }=\frac{\dot{\phi}^{2}}{2}+V\left( \phi \right) ,
\label{H} \\
2\dot{H}+3H^{2} &=&-p_{\phi }=-\frac{\dot{\phi}^{2}}{2}+V\left( \phi \right)
,  \label{H1}
\end{eqnarray}
where $\rho _{\phi }$ is the energy density, and $p_{\phi }$ is the pressure
due to the scalar field $\phi $, respectively, and the evolution equation
for the scalar field
\begin{equation}
\ddot{\phi}+3H\dot{\phi}+V^{\prime }\left( \phi \right) =0,  \label{phi}
\end{equation}%
where $H=\dot{a}/a>0$ is the Hubble expansion rate function. In the
following the overdot denotes the derivative with respect to the
time-coordinate $t$, and the prime denotes the derivative with respect to
the scalar field $\phi $, respectively. By adding Eqs.~(\ref{H}) and (\ref%
{H1}), we obtain the Riccati type equation satisfied by $H$, of the form
\begin{equation}
\dot{H}(t)=V(t)-3H^{2}(t).  \label{a3}
\end{equation}%
The deceleration parameter $q$, indicating the accelerating/decelerating
nature of the cosmological expansion, is defined as
\begin{equation}
q(t)=\frac{d}{dt}\left( \frac{1}{H}\right) -1=-\frac{\dot{H}}{H^{2}}-1=2-%
\frac{V(t)}{H^{2}(t)}.
\end{equation}

In the following Section, we will obtain several classes of exact solutions
of the Riccati Eq.~(\ref{a3}).


\section{Exact solutions of the Riccati evolution equation}
\label{sect3}


\subsection{The case $H(t)=\alpha V(t)+g(t)/3\beta $}


We assume that $H$ has the general form
\begin{equation}
H(t)=\alpha V(t)+\frac{1}{3\beta }g(t),
\end{equation}%
where $H(t)$, $V\left( t\right) $ and $g(t)\in C^{\infty }(I)$ are arbitrary
functions defined on a real time interval $I\subseteq \Re $ and $\alpha $,$%
\beta \in \Re $ are arbitrary constants. Hence Eq.~(\ref{a3}) can be written
as
\begin{equation}
\alpha \dot{V}+\frac{1}{3\beta }\dot{g}=V-3\alpha ^{2}V^{2}-2\frac{\alpha }{%
\beta }gV-\frac{1}{3\beta ^{2}}g^{2}.
\end{equation}


\subsubsection{The first class of solutions: $g\left( t\right) =\protect%
\beta /t$}


As a first example of an exact scalar field cosmological model we consider
that the arbitrary function $g\left( t\right) $ satisfies the differential
equation
\begin{equation}
\dot{g}=-\frac{1}{\beta }g^{2},
\end{equation}%
which gives for $g\left( t\right) $ the expression $g\left( t\right) =\beta
/t$, where without loss of generality, an arbitrary constant of integration
has been taken as zero. Then the scalar field potential satisfies the
Bernoulli differential equation
\begin{equation}
\dot{V}=\left( \frac{1}{\alpha }-\frac{2}{t}\right) V-3\alpha V^{2}.
\end{equation}

Therefore we have obtained the following theorem:

\textbf{Theorem 1}. If the self-interaction potential of a cosmological
scalar field has the time dependence given by
\begin{equation}
V(t)=\frac{e^{t/\alpha }}{t\left[ V_{0}t+3t\text{Ei}\left( \frac{t}{\alpha }%
\right) -3\alpha e^{t/\alpha }\right] },
\end{equation}%
where $Ei(z)=-\int_{-z}^{\infty }{e^{-t}dt/t}$ is the exponential integral
function \cite{PoZa}, and $V_{0}$ is an arbitrary constant of integration,
then the time variation of the Hubble function is obtained as
\begin{equation}
H(t)=\left[ 3t-\frac{9\alpha e^{t/\alpha }}{V_{0}+3\text{Ei}\left( \frac{t}{%
\alpha }\right) }\right] ^{-1}.
\end{equation}

Thus, the time variation of the scale factor is obtained as
\begin{equation}
a(t)=a_{0}\left\{ 3e^{t/\alpha }-\frac{t\left[ V_{0}+3\text{Ei}\left( \frac{t%
}{\alpha }\right) \right] }{\alpha }\right\} ^{1/3},
\end{equation}%
where $a_{0}$ is an arbitrary constant of integration. We obtain immediately
$\lim_{t\rightarrow 0}a(t)=3^{1/3}a_{0}$, showing that the obtained solution
is non-singular. The deceleration parameter is given by
\begin{equation}
q(t)=2-\frac{9e^{t/\alpha }\left[ V_{0}t+3t\text{Ei}\left( \frac{t}{\alpha }%
\right) -3\alpha e^{t/\alpha }\right] }{t\left[ V_{0}+3\text{Ei}\left( \frac{%
t}{\alpha }\right) \right] ^{2}}.
\end{equation}

\begin{centering}
\begin{figure*}[th]
\includegraphics[scale=0.67]{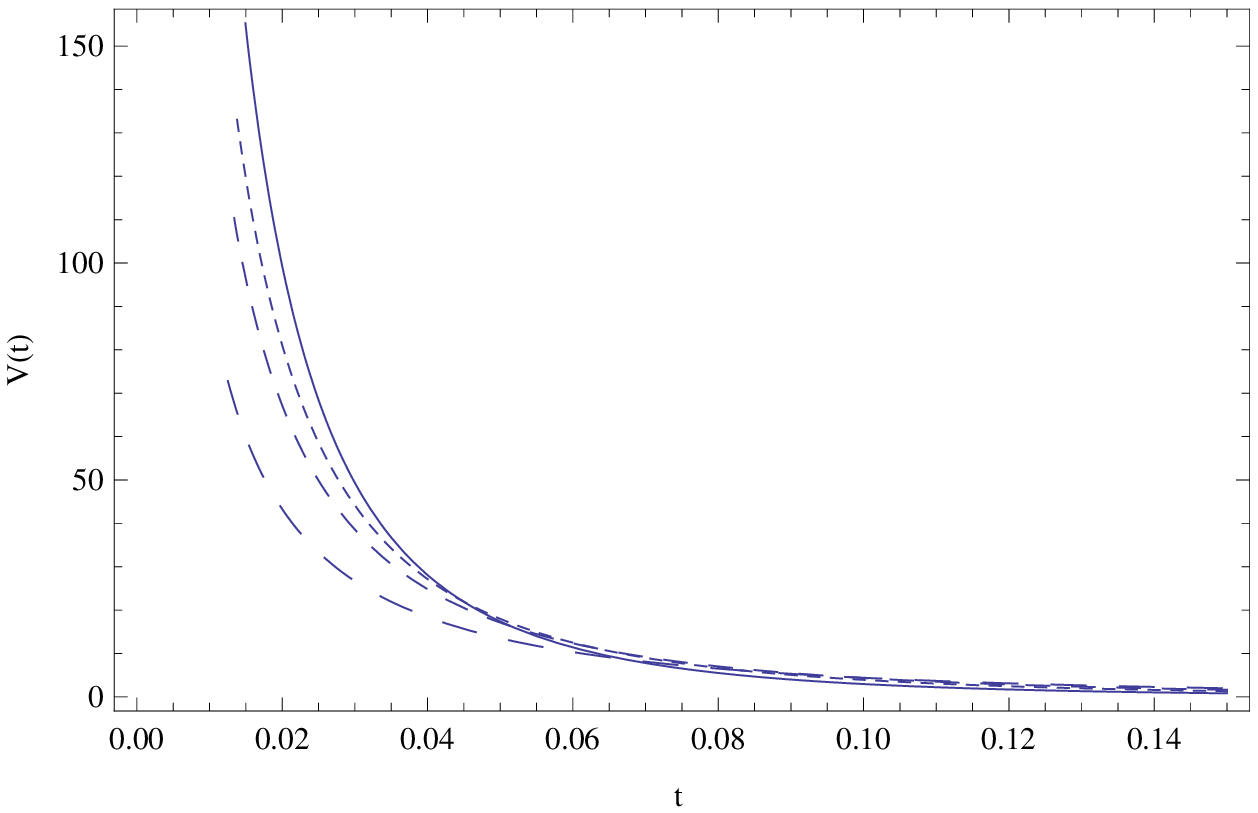}
\includegraphics[scale=0.67]{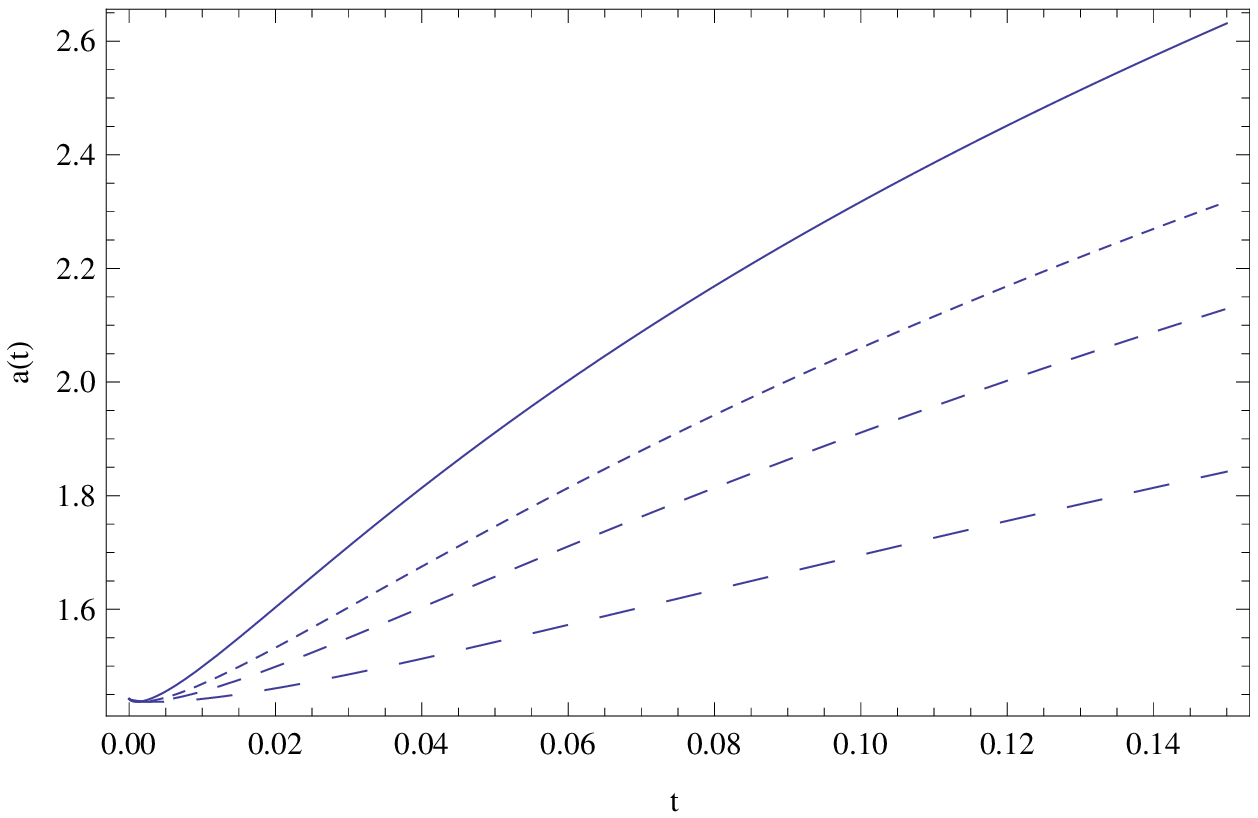}
\caption{Variation of the  scalar field potential (left plot), and of the scale factor (right plot), as a function of $t$ for the first solution of the Riccati equation, for different values of $\alpha $: $\alpha =-0.1 $ (solid curve), $\alpha =-0.15$ (dotted curve), $\alpha =-0.20$ (short dashed curve), and $\alpha =-0.35$ (dashed curve), respectively.  In all cases $V_0=12$, and $a_0=1$.  }\label{ricc1}
\end{figure*}
\end{centering}

\begin{centering}
\begin{figure*}[th]
\includegraphics[scale=0.61]{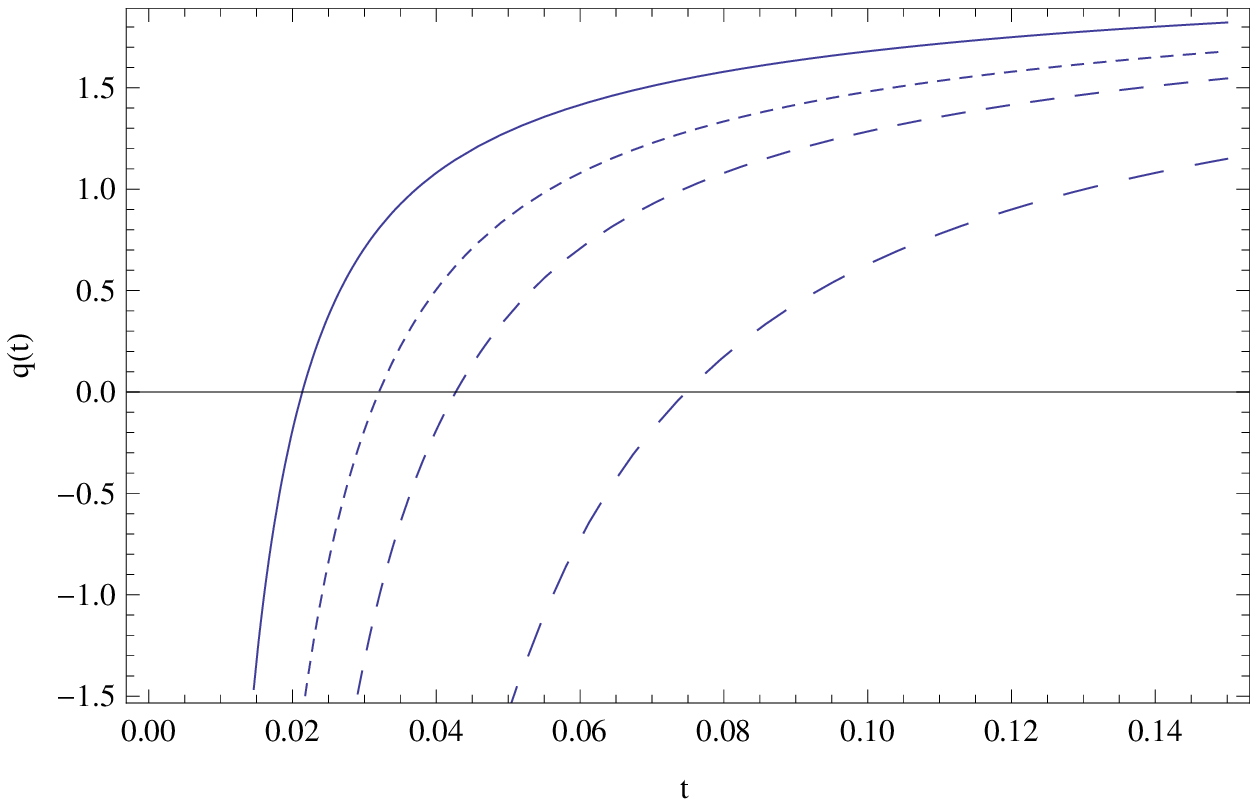}
\includegraphics[scale=0.73]{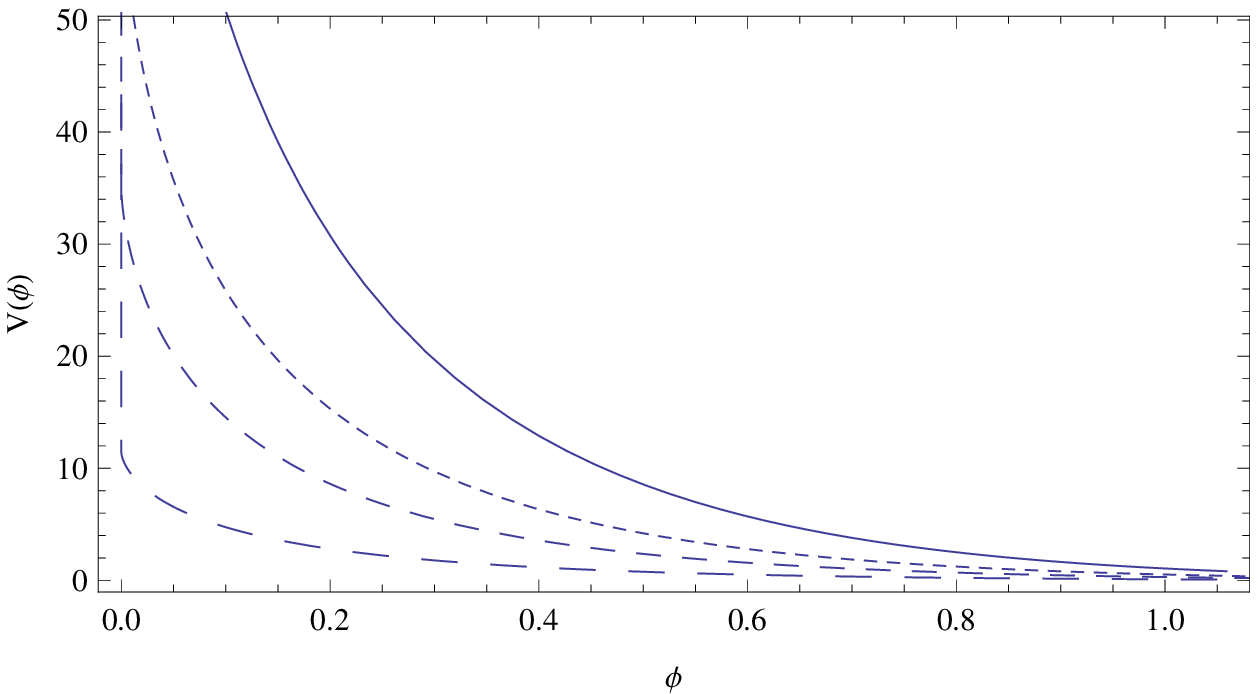}
\caption{Time variation of the deceleration parameter $q(t)$(left plot), and of $V(\phi)$ (right plot), for the first solution of the Riccati equation, for different values of  $\alpha $: $\alpha =-0.1 $ (solid curve), $\alpha =-0.15$ (dotted curve), $\alpha =-0.20$ (short dashed curve), and $\alpha =-0.35$ (dashed curve), respectively.  In all cases $V_0=12$. }\label{ricc4}
\end{figure*}
\end{centering}

The dynamics of the scalar field can be obtained from the  equation $%
\phi _{\pm }(t)-\phi _{0\pm }=\pm \sqrt{2}\int {\sqrt{3H^{2}(t)-V(t)}dt}$ as
\begin{widetext}
\be
\phi _{\pm }(t)-\phi _{0\pm }=\pm \sqrt{\frac{2}{3}}
\int ^t{\sqrt{\frac{9\xi \text{%
Ei}\left( \frac{\xi}{\alpha }\right) ^{2}-9\xi e^{\xi /\alpha }\text{Ei}%
\left( \frac{\xi}{\alpha }\right) +6V_{0}\xi \text{Ei}\left( \frac{\xi }{\alpha
}\right) +9\alpha e^{\frac{2\xi }{\alpha }}+V_{0}^{2}\xi-3V_{0}\xi e^{\xi/%
\alpha }}{\xi \left[ 3\xi \text{Ei}\left( \frac{\xi }{\alpha }\right)
-3\alpha e^{\xi /\alpha }+V_{0}\xi \right] ^{2}}}d\xi },
\ee
\end{widetext}
where $\phi _{0\pm }$ are arbitrary constants of integration. The time
variations of the scalar field potential, of the scale factor, of the
deceleration parameter, and the potential--scalar field dependence $V=V(\phi
)$ are presented, for different values of $\alpha $, and for a fixed value
of $V_{0}$, in Figs.~\ref{ricc1}-\ref{ricc4}.

\begin{centering}
\begin{figure}[h]
\includegraphics[scale=0.67]{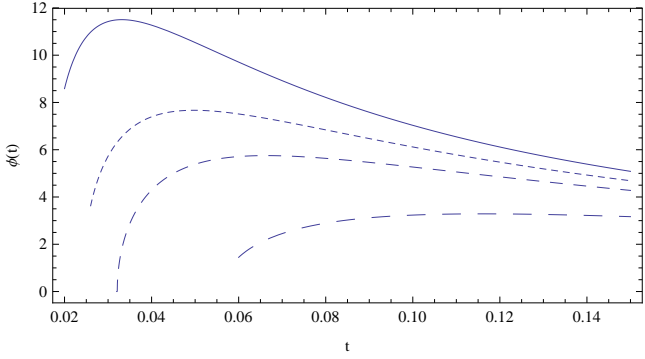}
\caption{Time dependence of the scalar field $\phi $  for the first solution of the Riccati equation for different values of $\alpha $:  $\alpha =-0.03 $ (solid curve), $\alpha =-0.08$ (dotted curve), $\alpha =-0.10$ (short dashed curve), and $\alpha =-0.12$ (dashed curve), respectively. In all cases $V_0=12$. }\label{ricc5}
\end{figure}
\end{centering}

The Universe starts its evolution from an accelerating phase, with $q\approx
-1.5$, but enters, after a short time interval, into a decelerating
expansionary phase, with $q>0$. Hence the present model can describe the early inflationary phase in the evolution of the Universe. However, the accelerated expansion is not of de Sitter type ($q=-1$). The scalar field potential tends rapidly to
zero, so that $\lim_{t\rightarrow \infty}V(t)=0$. The $V=V(\phi)$ dependence
of the potential on the scalar field, represented in the right plot of Fig.~%
\ref{ricc4}, cannot be obtained in an exact form. The time variation of the
scalar field $\phi $, represented in Fig.~\ref{ricc5}, shows that $\phi $ is
an increasing function of $t$.

In the small time limit, $t\rightarrow 0$, by taking into account the relation $\lim _{t\rightarrow 0}\;t\text{Ei}\left( t/\alpha \right)=0$, we obtain $V(t)\approx -1/3\alpha t$. In order to have a positive value of the potential we need to take $\alpha <0$. In the same limit of the small times and for $\alpha <0$, $H(t)$ can be obtained as
\bea
&&H(t)\approx \Bigg\{3 t \Bigg[\frac{6}{6 \ln |\alpha |-2
   V_0-6 \gamma }+\nonumber\\
  && \frac{36}{\left(6 \ln
   |\alpha |-2 V_0-6 \gamma \right)^2}+1\Bigg]+\frac{18 \alpha }{6 \ln |\alpha |-2 V_0-6 \gamma }\Bigg\}^{-1}\nonumber\\
   &&=\frac{1}{H_0t+\Lambda_0},
\eea
where $\gamma $ is Euler's constant. By assuming that $\Lambda _0$ can be neglected, we obtain for the scalar field $\phi \approx \lambda \ln t$, where $\lambda =\sqrt{3/H_0^2+1/3\alpha}$. Hence the functional relation between the scalar field potential and the potential can be approximated, in the early stages of the cosmological evolution, as $V(\phi)\approx V_0\exp(\phi /\lambda)$, which is the exponential potential that plays an important role in the study of the dynamics of the early Universe \cite{exppot,exppot2,exppot3,exppot4,exppot5}. Hence the present solutions of the gravitational field equations can be interpreted as representing a generalization of the scalar field models with exponential potential, to whom they reduce in the small time limit.


\subsubsection{The second class of solutions}


As a second example of an exact solution of the Riccati evolution equation
for the Hubble function we consider the case in which the potential
satisfies the Bernoulli differential equation
\begin{equation}
\dot{V}=\frac{1}{\alpha }V-3\alpha V^{2}.
\end{equation}%
Hence the time dependence of the scalar field potential can be obtained as
\begin{equation}
V(t)=\frac{e^{t/\alpha }}{V_{0}+3\alpha ^{2}e^{t/\alpha }},  \label{V2}
\end{equation}%
where $V_{0}$ is an arbitrary constant of integration. By assuming that at $%
t=0$, $V(0)=V_{\mathrm{in}}$, it follows that $V_{0}=1/V_{\mathrm{in}%
}-3\alpha ^{2}$. Then the function $g\left( t\right) $ must satisfy the
following Bernoulli differential equation,
\begin{equation}
\frac{1}{3}\dot{g}+2\alpha \frac{e^{t/\alpha }}{V_{0}+3\alpha
^{2}e^{t/\alpha }}g+\frac{1}{3\beta }g^{2}=0.
\end{equation}%
%
Therefore we have obtained the following

\textbf{Theorem 2}. If the time dependence of the scalar field potential is
given by Eq.~(\ref{V2}), then the Hubble function $H(t)$ is obtained as
\begin{widetext}
\be\label{H2}
H(t)=\frac{1}{3}\Bigg\{ \alpha +
\frac{1}{\left[ \beta
V_{0}^{2}g_{0}-\alpha \ln \left| V_{0}+3\alpha ^{2}e^{t/\alpha
}\right| +t\right] ^{-1}+3\left( \alpha /V_{0}\right) e^{\frac{t}{%
\alpha }}}\Bigg\} ^{-1},
\ee
\end{widetext}
where $g_0$ is an arbitrary constant of integration.
The scale factor is given by
\begin{eqnarray}
a(t)&=&a_{0}\Bigg[ 3\alpha ^{2}e^{t/\alpha }\left( \beta
V_{0}^{2}g_{0}+t\right) -\alpha \left( V_{0}+3\alpha ^{2}e^{t/\alpha
}\right) \times  \notag \\
&&\ln \left| V_{0}+3\alpha ^{2}e^{t/\alpha }\right| +V_{0}\left( \alpha
+\beta V_{0}^{2}g_{0}+t\right) \Bigg]^{\frac{1}{3}} ,
\end{eqnarray}
where we have denoted $a_{0}=1/V_{0}$. It is interesting to note that the
scale factor $a(t)$ is non-singular at $t=0$, and
\begin{eqnarray}
\lim_{t\rightarrow 0}a(t)&=&a_{0}\Bigg[V_{0}\left( \alpha +3\alpha ^{2}\beta
V_{0}g_{0}+\beta V_{0}^{2}g_{0}\right)   \notag \\
&&-\alpha \left( 3\alpha ^{2}+V_{0}\right) \ln \left| 3\alpha
^{2}+V_{0}\right|\Bigg]^{1/3} .
\end{eqnarray}

The deceleration parameter in this model is given by
%
\begin{eqnarray}
&&q(t)=-1
   \nonumber \\
&& \hspace{-0.9cm} -\frac{3\left[ \frac{3e^{t/\alpha }}{V_{0}}-\frac{V_{0}}{\left(
3\alpha ^{2}e^{t/\alpha }+V_{0}\right) \left( \beta
g_{0}V_{0}^{2}-\alpha \ln \left| 3\alpha ^{2}e^{t/\alpha
}+V_{0}\right| +t\right) {}^{2}}\right] }{\left[ \frac{1}{\beta
g_{0}V_{0}^{2}-\alpha \ln \left| 3\alpha^{2}e^{t/\alpha
}+V_{0}\right| +t}+\frac{3\alpha e^{t/\alpha }}{V_{0}}\right]
^{2}}.
\end{eqnarray}

The time variation of the scalar field can be obtained from the equation $\phi _{\pm}(t)-\phi _{0\pm}=\pm \sqrt{2}\int ^t{\sqrt{3H^2(\xi)-V(\xi )}d\xi}$, and the dependence of the potential on the scalar field is obtained in a parametric form.

In the limit of small times, from Eq.~(\ref{V2}) it follows that $V(t)\approx 1/\left(V_0+3\alpha ^2\right)=V_{in}={\rm constant}$, while $\lim_{t\rightarrow 0}H(t)\approx \left\{ \alpha +\left[ \left[ \beta
V_{0}^{2}g_{0}-\alpha \ln \left\vert V_{0}+3\alpha ^{2}\right\vert \right]
^{-1}+3\left( \alpha /V_{0}\right) \right] \right\} ^{-1}/3=H_0={\rm constant}$. The deceleration parameter has an initial value given by
\be
\lim _{t\rightarrow 0}q(t)=-\frac{9 \left[\frac{\alpha ^2}{\left(3 \alpha ^2+V_0\right) \left(\beta  g_0 V_0^2-\alpha  \ln \left|3 \alpha
   ^2+V_0\right|\right)^2}+\frac{1}{V_0}\right]}{\left(\frac{1}{\beta  g_0 V_0^2-\alpha  \ln \left|3 \alpha ^2+V_0\right|}+\frac{3 \alpha
   }{V_0
   }\right)^2}-1,
   \ee
and, depending on the numerical values of the free parameters of the model, a large number of initial states can be constructed.
 During the initial stages of the expansion, the scalar field is obtained as $\phi _{\pm}(t)-\phi _{0\pm}=\pm \sqrt{3H_0^2-V_{in}}t$, and shows a linear increase in time during the early phases of the evolution of the Universe.

In the limit $t\rightarrow \infty $, the scalar field potential tends to a
constant, $\lim_{t\rightarrow \infty }V(t)=1/3\alpha ^{2}$, while $%
\lim_{t\rightarrow \infty }H(t)=1/3\alpha $, and $\lim_{t\rightarrow \infty
}q(t)=-1$, respectively. Therefore for this scalar field potential the
Universe ends in a de Sitter-type exponential expanding phase. During its entire evolution, the Universe remains in an accelerating phase, with the deceleration parameter having negative values $q \leq-1$. Such a scalar field model may be used for the description of the late evolutionary stages of the expansion of the Universe, and could represent an effective dark energy model. Alternatively, from a cosmological point of view this model can be interpreted as describing an eternally inflating Universe.


\subsection{Exact solution of the field equations for $V\left( t\right)
=f_{1}\left( t\right) +3\left[ \protect\int^{t}f_{1}\left( \protect\xi %
\right) d\protect\xi +V_{1}\right] ^{2}$}


We assume that the potential $V\left( t\right) $ satisfies the integral
condition%
\begin{equation}
V\left( t\right) =f_{1}\left( t\right) +3\left[ \int^{t}f_{1}\left( \xi
\right) d\xi +V_{1}\right] ^{2},  \label{n6}
\end{equation}%
where we have introduced a new arbitrary function $f_{1}(t)\in C^{\infty
}(I) $ defined on a real interval $I\subseteq \Re $ and $V_{1}\in \Re $ is
an arbitrary constant. By inserting Eq.~(\ref{n6}) into Eq.~(\ref{a3}), the
latter takes the form
\begin{equation}
\frac{dH}{dt}=f_{1}\left( t\right) +3\left[ \int^{t}f_{1}\left( \xi \right)
d\xi +V_{1}\right] ^{2}-3H^{2}.  \label{n10}
\end{equation}

Therefore we obtain the following:

\textbf{Theorem 3}. If the scalar field potential $V(t)$ satisfies the
integral condition (\ref{n6}), then the general solution of the Riccati
equation (\ref{a3}) is given by
\begin{eqnarray}
H\left( t\right) &=&\frac{e^{-6V_{1}t-6\int^{t}\int^{\psi }f_{1}\left( \xi
\right) d\xi d\psi }}{C_{1}+3\int^{t}e^{-6V_{1}\eta -6\int^{\eta }\int^{\psi
}f_{1}\left( \xi \right) d\xi d\psi }d\eta }  \notag \\
&&+\int^{t}f_{1}\left( \xi \right) d\xi +V_{1},  \label{n11}
\end{eqnarray}
where $C_{1}$ is an arbitrary constant of integration.

Equation~(\ref{n11}) can be immediately integrated to give the scale factor
in the form
\begin{eqnarray}
a\left( t\right) &=&a_{0}e^{V_{1}t+\int^{t}\int^{\zeta }f_{1}\left( \xi
\right) d\xi d\zeta } \times  \notag \\
&& \left[ C_{1}+3\int^{t }e^{-6V_{1}\eta -6\int^{\eta }\int^{\psi
}f_{1}\left( \xi \right) d\xi d\psi }d\eta \right] ^{1/3},
\end{eqnarray}
where $a_{0}$ is an arbitrary constant of integration. The deceleration
parameter $q$ takes the form 
%
\begin{eqnarray}
q\left( t\right) &=&2- \left[ f_{1}\left( t\right) +3\left(
\int^{t}f_{1}\left( \xi \right) d\xi +V_{1}\right) ^{2} \right]\times  \notag
\\
&&\Bigg[ \frac{e^{-6V_{1}t-6\int^{t}\int^{\psi }f_{1}\left( \xi \right) d\xi
d\psi }}{C_{1}+3\int^{t}e^{-6V_{1}\eta -6\int^{\eta }\int^{\psi }f_{1}\left(
\xi \right) d\xi d\psi }d\eta }  \notag \\
&&+\int^{t}f_{1}\left( \xi \right) d\xi +V_{1}\Bigg] ^{-2}.  \label{n13}
\end{eqnarray}

The scalar field $\phi \left( t\right) $ can be written as
\begin{eqnarray}
\phi _{\pm }\left( t\right) &=&\phi _{0\pm }\pm \sqrt{2}\int^{t}\Bigg\{ %
-f_{1}(\zeta )-3\left[ \int^{\zeta }f_{1}(\xi )d\xi +V_{1}\right] ^{2}
\notag \\
&& +3\Bigg[ \frac{e^{-6V_{1}\zeta -6\int^{\zeta }\int^{\psi }f_{1}\left( \xi
\right) d\xi d\psi }}{C_{1}+3\int^{\zeta }e^{-6V_{1}\eta -6\int^{\eta
}\int^{\psi }f_{1}\left( \xi \right) d\xi d\psi }d\eta }  \notag \\
&& +\int^{\zeta }f_{1}\left( \xi \right) d\xi +V_{1}\Bigg] ^{2} \Bigg\}^{%
\frac{1}{2}}d\zeta ,  \label{n14}
\end{eqnarray}%
where $\phi _{0\pm }$ are arbitrary constants of integration.

Note that the physical behavior of the Universe is determined by the choice
of the arbitrary function $f_{1}\left( t\right) $. The potential $V\left(
\phi \right) $ can be uniquely determined in a parametric form from Eqs.~(%
\ref{n6}) and (\ref{n14}), with the time $t$ taken as a parameter.\newline


As a simple application of the integrability condition of the gravitational
field equations given by \textbf{Theorem 3} we consider the particular case
for which $f_{1}(t)=f_{0}=\mathrm{constant}>0$. Moreover, for simplicity, we
also assume $V_{1}=0$. Therefore, in this model, the time dependence of the
scalar field potential is given by
\begin{equation}
V(t)=3f_{0}^{2}t^{2}+f_{0}.  \label{pot11}
\end{equation}%
For the Hubble function we obtain the expression
\begin{equation}
H(t)=\frac{2\sqrt{f_{0}}e^{-3f_{0}t^{2}}}{2H_{1}\sqrt{f_{0}}+\sqrt{3\pi }%
\text{erf}\left( \sqrt{3f_{0}}t\right) }+f_{0}t,
\end{equation}%
where $H_{1}$ is an arbitrary constant of integration, and $\mathrm{erf}%
(z)=\left( 2/\sqrt{\pi }\right) \int_{0}^{z}{\exp \left( -t^{2}\right) dt}$
is the error function, giving the integral of the Gaussian distribution \cite%
{PoZa}. The scale factor $a(t)$ can be obtained as
\begin{equation}
a(t)=a_{0}e^{\frac{f_{0}t^{2}}{2}}\left[ H_{1}+\frac{1}{2}\sqrt{\frac{3\pi }{%
f_{0}}}\text{erf}\left( \sqrt{3f_{0}}t\right) \right] ^{1/3}.
\end{equation}

The deceleration parameter is given by
\begin{eqnarray}
&&q(t)=-1+\frac{1}{f_{0}t^{2}}\times  \notag \\
&&\Bigg\{\frac{4\left( 3f_{0}t^{2}+1\right) }{\sqrt{f_{0}}te^{3f_{0}t^{2}}%
\left[ 2H_{1}\sqrt{f_{0}}+\sqrt{3\pi }\text{erf}\left( \sqrt{3f_{0}}t\right) %
\right] +2}-  \notag \\
&&\frac{4\left( 3f_{0}t^{2}+1\right) }{\left\{ \sqrt{f_{0}}te^{3f_{0}t^{2}}%
\left[ 2H_{1}\sqrt{f_{0}}+\sqrt{3\pi }\text{erf}\left( \sqrt{3f_{0}}t\right) %
\right] +2\right\} ^{2}}-1\Bigg\}.  \notag \\
\end{eqnarray}

The time dependence of the scalar field can be obtained in an integral form
as
\begin{eqnarray}  \label{phi11}
\phi (t) &=&\phi _{0}+\sqrt{2}\int^{t}\Big[-3f_{0}^{2}\zeta ^{2}-3f_{0}\times
\notag \\
&&\hspace{-1cm}\left( \frac{2\sqrt{f_{0}}e^{-3f_{0}\zeta ^{2}}}{2H_{1}\sqrt{%
f_{0}}+\sqrt{3\pi }\text{erf}\left( \sqrt{3f_{0}}\zeta \right) }+f_{0}\zeta
\right) ^{2}\Bigg]^{1/2}\;d\zeta .
\end{eqnarray}

Eqs.~(\ref{pot11}) and (\ref{phi11}) give the functional dependence of the
scalar field potential $V$ of the scalar field $\phi $ in a parametric form,
with $t$ taken as parameter.

In the small time limit the Hubble parameter can be approximated as
\be
H(t)\approx  \left(f_0-\frac{3}{H_1^2}\right)t+\frac{1}{H_1},
\ee
while in the same order of approximation the expression $\sqrt{3H^2-V}$ can be obtained as
\be
\sqrt{3H^2-V}\approx \frac{\sqrt{3-f_0 H_1^2} (H_1-3 t)}{H_1^2}.
\ee
Therefore in the small time limit for the scalar field evolution we find
\be
\phi (t)\approx \frac{\sqrt{3-f_0 H_1^2} \left(H_1 t-\frac{3
   t^2}{2}\right)}{H_1^2}.
\ee 

In the first order we obtain the time-scalar field dependence as
\be
t(\phi)\approx \frac{H_1 \phi }{\sqrt{3-f_0 H_1^2}},
\ee
which gives the scalar field potential as a function of $\phi $ in the form
\be
V(\phi)\approx \frac{3 f_0^2 H_1^2 }{3-f_0 H_1^2}\phi ^2+f_0.
\ee
Quadratic potentials have been extensively investigated in the recent literature \cite{quad,quad2,quad3,quad4}, and they allow to recover the connection with particle physics. The effective mass of the scalar field is given by $m_{\phi}=3 f_0^2 H_1^2/2\left(3-f_0 H_1^2\right)$. Moreover, the constant term in the potential naturally generates a cosmological constant. The early time evolution of the deceleration parameter can be approximated as
\bea
q(t)&\approx& -3 f_0  \left(f_0^2 H_1^4-3 f_0 H_1^2+3\right)t^2+\nonumber\\
&&2f_0 H_1  \left(f_0 H_1^2-3\right)t-f_0
   H_1^2+2.
\eea 
If $f_0H_1^2>2$, the Universe starts its evolution from an accelerating phase. Hence the present model can describe a generalized effective power law type  scalar field potential, which in the small time limit reduces to the quadratic potential.

\subsection{Exact solution of the field equations for $V_{\pm}\left(
t\right) =3f_{2}\left( t\right) \pm \frac{d}{dt}\protect\sqrt{f_{2}\left(
t\right) }$}


We assume that the potential $V_{\pm }\left( t\right) $ satisfy the
differential condition%
\begin{equation}
V_{\pm }\left( t\right) =3f_{2}\left( t\right) \pm \frac{d}{dt}\sqrt{%
f_{2}\left( t\right) },  \label{v6}
\end{equation}%
where we have introduced a new arbitrary function $f_{2}(t)\in C^{\infty }(I)
$ defined on a real interval $I\subseteq \Re $. By inserting Eq.~(\ref{v6})
into Eq.~(\ref{a3}), the latter takes the form
\begin{equation}
\frac{dH_{\pm }}{dt}=3f_{2}\left( t\right) \pm \frac{d}{dt}\sqrt{f_{2}\left(
t\right) }-3H_{\pm }^{2}.  \label{v10}
\end{equation}

Therefore we have obtained the following:

\textbf{Theorem 4}. If the scalar field potential $V(t)$ satisfies the
differential condition (\ref{v6}), then the general solutions of the Riccati
Eq.~(\ref{a3}) are given by
\begin{equation}
H_{\pm }\left( t\right) =\frac{e^{\mp 6\int^{t}\sqrt{f_{2}\left( \psi
\right) }d\psi }}{C_{1\pm }+3\int^{t}e^{\mp 6\int^{\eta }\sqrt{f_{2}\left(
\psi \right) }d\psi }d\eta }\pm \sqrt{f_{2}\left( t\right) },  \label{v11}
\end{equation}%
where $C_{1\pm }$ are arbitrary constants of integration. Equation~(\ref{v11}%
) can be integrated to give the scale factor in the form
\begin{eqnarray}
a_{\pm }\left( t\right)  &=&a_{0\pm }e^{\pm \int^{t}\sqrt{f_{2}\left( \zeta
\right) }d\zeta }\times   \notag \\
&&\left[ C_{1\pm }+3\int^{t}e^{\mp 6\int^{\eta }\sqrt{f_{2}\left( \psi
\right) }d\psi }d\eta \right] ^{1/3},
\end{eqnarray}%
where $a_{0\pm }$ are arbitrary constants of integration.

With the help of Eqs.~(\ref{a3}), (\ref{v6}), and (\ref{v11}), respectively,
the deceleration parameter $q$ can be written as
\begin{equation}
q_{\pm }\left( t\right) =2-\frac{3f_{2}\left( t\right) \pm \frac{d}{dt}\sqrt{%
f_{2}\left( t\right) }}{\left[ \frac{e^{\mp 6\int^{t}\sqrt{f_{2}\left( \psi
\right) }d\psi }}{C_{1\pm }+3\int^{\zeta }e^{\mp 6\int^{\eta }\sqrt{%
f_{2}\left( \psi \right) }d\psi }d\eta }\pm \sqrt{f_{2}\left( t\right) }%
\right] ^{2}}.  \label{v13}
\end{equation}

The scalar field $\phi \left( t\right) $ can be written as
\begin{eqnarray}  \label{v14}
&&\phi _{\pm }\left( t\right) =\phi _{0\pm }\pm \sqrt{2}\int^{t}\Big\{%
-3f_{2}\left( \zeta \right) \mp \frac{d}{d\zeta }\sqrt{f_{2}\left( \zeta
\right) }+  \notag \\
&&3\left[ \frac{e^{\mp 6\int^{\zeta }\sqrt{f_{2}\left( \psi \right) }d\psi }%
}{C_{1\pm }+3\int^{\zeta }e^{\mp 6\int^{\eta }\sqrt{f_{2}\left( \psi \right)
}d\psi }d\eta }\pm \sqrt{f_{2}\left( \zeta \right) }\right] ^{2}\Bigg\}^{%
\frac{1}{2}}\,d\zeta .  \notag \\
\end{eqnarray}%
where $\phi _{0\pm }$ are arbitrary constants of integration.\newline


As an example of the application of \textbf{Theorem 4} we consider the case
in which the function $f_{2}(t)$ has the form $f_{2}(t)=f_{02}/t^{2}$, with $%
f_{02}=\mathrm{constant}>0$. In this case the scalar field potential takes
the form
\begin{equation}
V_{\pm }(t)=\frac{V_{0\pm }}{t^{2}},  \label{jj}
\end{equation}%
where for simplicity, we have introduced the arbitrary constants $V_{0\pm }$
$\ $defined as $V_{0\pm }=3f_{02}\mp \sqrt{f_{02}}$. The Hubble function can
then be obtained immediately either from Eqs.~(\ref{v6}), (\ref{v10},( \ref{jj})
or from Eq. (\ref{v11}) as
\begin{equation}
H_{\pm }(t)=\frac{1}{6t}\left[ 1+V_{1\pm }\left( 1-\frac{2H_{1\pm }}{H_{1\pm
}+t^{V_{1\pm }}}\right) \right] \mathbf{,}
\end{equation}
where $H_{1\pm }$ are arbitrary constants of integration, and $V_{1\pm }=%
\sqrt{12V_{0\pm }+1}$. For the scale factor we obtain
\begin{equation}
a_{\pm }(t)=a_{0\pm }t^{\left( 1-V_{1\pm }\right) /6}\left( H_{1\pm
}+t^{V_{1\pm }}\right) ^{1/3}\mathbf{,}
\end{equation}
where $a_{0\pm }$ are arbitrary constants of integration. 
The deceleration parameter is given by
\begin{equation}
q_{\pm }(t)=\frac{4\left[ -30H_{1\pm }V_{0\pm }t^{V_{1\pm }}-H_{1\pm
}^{2}q_{1\pm }+q_{2\pm }t^{2V_{1\pm }}\right] }{\left[ H_{1\pm }\left(
1-V_{1\pm }\right) +\left( V_{1\pm }+1\right) t^{V_{1\pm }}\right] ^{2}},
\end{equation}%
where $q_{1\pm }=3V_{0\pm }+V_{1\pm }-1$, and $q_{2\pm }=-3V_{0\pm }+V_{1\pm
}+1$. The scalar field $\phi \left( t\right) $ can be written as
\begin{eqnarray}  \label{qw}
&&\phi _{\pm }\left( t\right) =\phi _{0\pm }\pm \frac{1}{\sqrt{6}}\times
\notag \\
&&\int^{t}\frac{1}{\zeta }\sqrt{\left[ V_{1\pm }\left( \frac{2H_{1\pm }}{%
H_{1\pm }+\zeta ^{V_{1\pm }}}-1\right) -1\right] ^{2}-12V_{0\pm }}\,d\zeta .
\notag \\
\end{eqnarray}

Note that the integral on the right hand side of Eq.~(\ref{qw}) can be
evaluated exactly with a very complicated expression. However in order to
have a concise representation we keep the integral form of the scalar field
here. In the following we restrict our analysis to the case $V_{0+}=-1/12$,
corresponding to the value $f_{02}=1/36$. In this case we obtain a complete
particular solution of the gravitational field equations describing the time
evolution of the flat FRW Universe with the self interaction potential $%
V_{+}(\phi )$, given by
\begin{eqnarray}
&&H_{+}(t)=\frac{1}{6t},a_{+}(t)=a_{0+}t^{1/6},q_{+}=5,  \notag \\
&&\phi _{+}(t)=\phi _{0+}+\frac{\ln \left| t\right| }{\sqrt{3}},V_{+}(\phi
)=V_{2+}e^{-2\sqrt{3}\phi },
\end{eqnarray}
where for simplicity, we have denoted $V_{2+}=-e^{2\sqrt{3}\phi _{0+}}/12$.
Thus we have regained the simple power law solution for the cosmological
model with the potential expressed as the exponential function of the scalar
field \cite{pow1}. This solution represents a decelerating cosmology, with $q>0$, and it may be useful for the description of the post-inflationary decelerating phase of the early Universe, or during the reheating period. 


\section{Discussions and final remarks}

\label{sect7}

In the present paper, we have shown that the time evolution and dynamics of
the Hubble function in scalar field cosmologies can be formulated in terms
of a simple first order Riccati type equation, with the cosmological
dynamics entirely determined by the time variation of the scalar field
potential. This equation immediately leads to the identification of some
classes of scalar field potentials for which the field equations can be
solved exactly, and it allows the formulation of very general integrability
conditions. We have obtained the complete solution of the gravitational
field equations describing the time evolution of the flat FRW Universe in
the presence of the scalar field $\phi \left( t\right) $ for four functional
forms of the self-interaction potential $V$. The first two solutions are
obtained for fixed forms of the scalar field potential, while in the last
two solutions the form of the potential is arbitrary, and determined by a
general integrability condition. The integrability conditions determine the
allowed form of the scalar field self-interaction potential in terms of some
arbitrary time dependent functions $f_{1}\left( t\right) $ and $f_{2}\left(
t\right) $, respectively, thus leading to the possibility of constructing
very general solutions of the field equations, and to reconstruct easily the
Hubble function, once the evolution of the potential is given.

In conclusion, we have obtained several exact solutions of the gravitational
field equations in the presence of a scalar field. In order to obtain a
deeper physical understanding of the solutions  comparisons with the
observational data are necessary. Work under these lines is presently
underway, and the results will be presented in a future publication.


\section*{Acknowledgments}


We would like to thank the anonymous referee for comments and suggestions that helped us to significantly improve our manuscript. FSNL is supported by a Funda\c{c}\~{a}o para a Ci\^{e}ncia e Tecnologia
Investigador FCT Research contract, with reference IF/00859/2012, funded by
FCT/MCTES (Portugal). FSNL also acknowledges financial support of the Funda%
\c{c}\~{a}o para a Ci\^{e}ncia e Tecnologia through the grants
CERN/FP/123615/2011 and CERN/FP/123618/2011. MKM acknowledges financial
support from the Vocational Training Council, Hong Kong.


\end{document}